\documentclass[prl,aps,twocolumn,showpacs,superscriptaddress,floatfix]{revtex4}
\usepackage{graphicx}
\usepackage{wasysym}
\usepackage{bm}

\newcommand{\beq}{\begin{equation}}
\newcommand{\eeq}{\end{equation}}
\newcommand{\bea}{\begin{eqnarray}}
\newcommand{\eea}{\end{eqnarray}}

\begin{document}

\title{Classical antiferromagnet on a hyperkagome lattice}

\author{John M. Hopkinson}
\affiliation{Department of Physics, University of Toronto, Toronto,
Ontario M5S 1A7, Canada}
\author{Sergei V. Isakov}
\affiliation{Department of Physics, University of Toronto, Toronto,
Ontario M5S 1A7, Canada}
\author{Hae-Young Kee}
\affiliation{Department of Physics, University of Toronto, Toronto,
Ontario M5S 1A7, Canada}
\author{Yong Baek Kim}
\affiliation{Department of Physics, University of Toronto, Toronto,
Ontario M5S 1A7, Canada}
\affiliation{Department of Physics, University of California, Berkeley,
California 94720}

\date{\today}

\begin{abstract}

Motivated by recent experiments on Na$_4$Ir$_3$O$_8$
[Y. Okamoto, M. Nohara, H. Aruga-Katori, and H. Takagi, arXiv:0705.2821
(unpublished)], 
we study the classical antiferromagnet on a frustrated three-dimensional 
lattice obtained by selectively removing one of four sites 
in each tetrahedron of the pyrochlore lattice. This ``hyperkagome'' lattice
consists of corner-sharing triangles. We present the results of 
large-$N$ mean field theory and Monte Carlo computations on 
$O(N)$ classical spin models. 
It is found that the classical ground states are highly degenerate.
Nonetheless a nematic order emerges at low temperatures in the
Heisenberg model ($N=3$) via ``order by disorder'', representing
the dominance of coplanar spin configurations. 
Implications for ongoing experiments are discussed.

\end{abstract}


\maketitle

Antiferromagnets on geometrically frustrated lattices often possess 
macroscopically degenerate classical ground states that satisfy peculiar local
constraints imposed by the underlying lattice structure \cite{moessner01}.
Such highly degenerate systems are extremely sensitive to thermal and 
quantum fluctuations, and thereby intriguing classical and quantum ground
states may emerge via ``order by disorder'' \cite{villain}.  On the other hand,
systems may remain disordered even at zero temperature \cite{anderson}.
These paramagnetic
states are called spin liquid phases and their classical and quantum
varieties have been recent subjects of intensive theoretical and experimental
research activities \cite{moessner01}. Excitement in such spin systems \cite{nigas} has also led to developments in mesoscopics \cite{wang}, optical lattices \cite{lowen}, and quantum coherence and computing \cite{osborne}. 

Among several examples of two and three-dimensional frustrated magnets,
the kagome and pyrochlore lattices have obtained particular attention because
a relatively large number of materials with the magnetic ions sitting on these 
lattice structures are available \cite{moessner01}.
Both of these lattices are corner-sharing 
structures of a basic unit; the triangle and tetrahedron respectively.
Despite this
similarity, the classical Heisenberg magnet orders on the kagome lattice 
\cite{chalker92,kagome} while
it remains disordered on the pyrochlore lattice \cite{moessner98}.
The nature of the spin-1/2
quantum Heisenberg magnets on these lattices has not been settled
and remains an important open problem \cite{sachdev,henley}.
On the other hand, spin-1/2
systems are rare on these lattices and other degrees of freedom such as
lattice distortions may play an important additional role. As a result,
direct experimental tests on spin-1/2 quantum magnets have 
been difficult to realize.

\begin{figure}[t]
\includegraphics[width=2.4in]{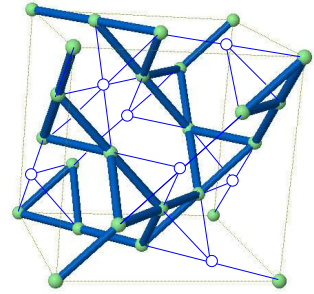}
\caption{ (color online). The hyperkagome lattice. 
The thin lines show the underlying pyrochlore lattice.}
\label{lattice}
\end{figure}

In this context, the recent experiments on Na$_4$Ir$_3$O$_8$ \cite{takagi}
may provide an important clue on these issues, albeit in a different
three-dimensional frustrated lattice. Here Ir$^{4+}$ carries spin-1/2
as the five $d$-electrons form a low spin state in the $t_{2g}$ level.
The Ir and Na ions together occupy the sites of the pyrochlore lattice
such that only three of the four sites of each tetrahedron are
occupied by Ir. The resulting lattice of magnetic Ir is
a network of corner-sharing triangles as shown in Fig.~\ref{lattice}, 
where each triangle is derived from different faces of the tetrahedra.
In analogy to the kagome lattice in two-dimensions, 
it is called the hyperkagome lattice.
Even though the Curie-Weiss temperature is large,
$\theta_W = - 650 K$, the susceptibility and specific heat
show no sign of magnetic ordering, nor lattice distortion,
down to $T \sim |\theta_W|/200$ \cite{takagi}; 
suggesting that it may be a spin liquid down to low temperatures.

In this paper, we study the classical antiferromagnet on the
hyperkagome lattice. Such investigations not only reveal the 
behavior of the antiferromagnet in the classical regime, but also
provide an important starting ground for the understanding of
quantum fluctuation effects.  We first study the large-$N$ limit of the 
$O(N)$ vector spin model at zero temperature and 
compute the spin-spin correlation function
in the large-$N$ mean field theory \cite{garcan,isakov04}.
It is found that there exist 
macroscopically degenerate ground states. 

Then we perform large-scale Monte Carlo computations on the 
Ising ($N=1$) and the Heisenberg ($N=3$) models.
The Heisenberg model (with exchange coupling $J$) remains disordered down to 
quite low temperatures, exhibiting very similar spin-spin correlations to those
of the large-$N$ model. 
These correlations show a characteristic dipolar structure in the reciprocal space, 
which can be explained by a mapping to a gauge theory \cite{isakov04,dipolar}.
On the other hand, the spin correlations in the Ising ($N=1$) model 
turn out to be quite different.

Most interestingly, a first order transition to a long range nematic order is 
observed in the Heisenberg model at a finite temperature. 
As explained below, this nematic order emerges via an ``order by disorder''
effect and represents the dominance of coplanar spin configurations. 
In the disordered phase no evidence of magnetic ordering is found 
while our numerical data cannot
definitely confirm the presence/absence of magnetic order
at temperatures below the onset of nematic order. 
We have also investigated the effect of an external 
magnetic field, $h$,
and found that a collinear order is chosen when $h = 2J$ in analogy
to a similar study on the kagome lattice \cite{zhit}.

{\it The lattice and local constraints}.---The hyperkagome lattice,
relevant to Na$_4$Ir$_3$O$_8$, 
can be represented by the simple cubic lattice with a 
twelve-site basis, as shown in Fig.~\ref{lattice}. This lattice is also
a three-dimensional network of corner-sharing triangles.
The model for 
the classical nearest-neighbor antiferromagnet on the 
hyperkagome lattice can be written as
\begin{equation}
H = J \sum_{\langle ij \rangle} {\bf S}_i \cdot {\bf S}_j 
= {J \over 2}  \sum_{\Delta} ({\bf S}_{\Delta})^2 +  {\rm constant},
\end{equation}
where $J > 0$, $\langle i,j \rangle$ represents the sum over
the nearest-neighbors, and ${\bf S}_i = (S^1_i,...,S^N_i)$ 
are $N$-component spins of fixed length $N$.
${\bf S}_{\Delta} = \sum_{i \epsilon \Delta} {\bf S}_i$ is the vectorial
sum of the spins in each triangle and $\sum_{\Delta}$ represents
the sum over all triangles. For $N \ge 2$, the classical ground state 
satisfies $S_{\Delta}=0$ for every triangle while the constraint
on the Ising model is $S_{\Delta} = \pm 1$. Thus, from the outset,
one may expect that the physics of the larger-$N$ models would be
different from the Ising case. This is different from the
antiferromagnet
on the pyrochlore lattice where the Ising and Heisenberg models satisfy 
the same constraint. 
\begin{figure}[t]
\includegraphics[width=1.4in]{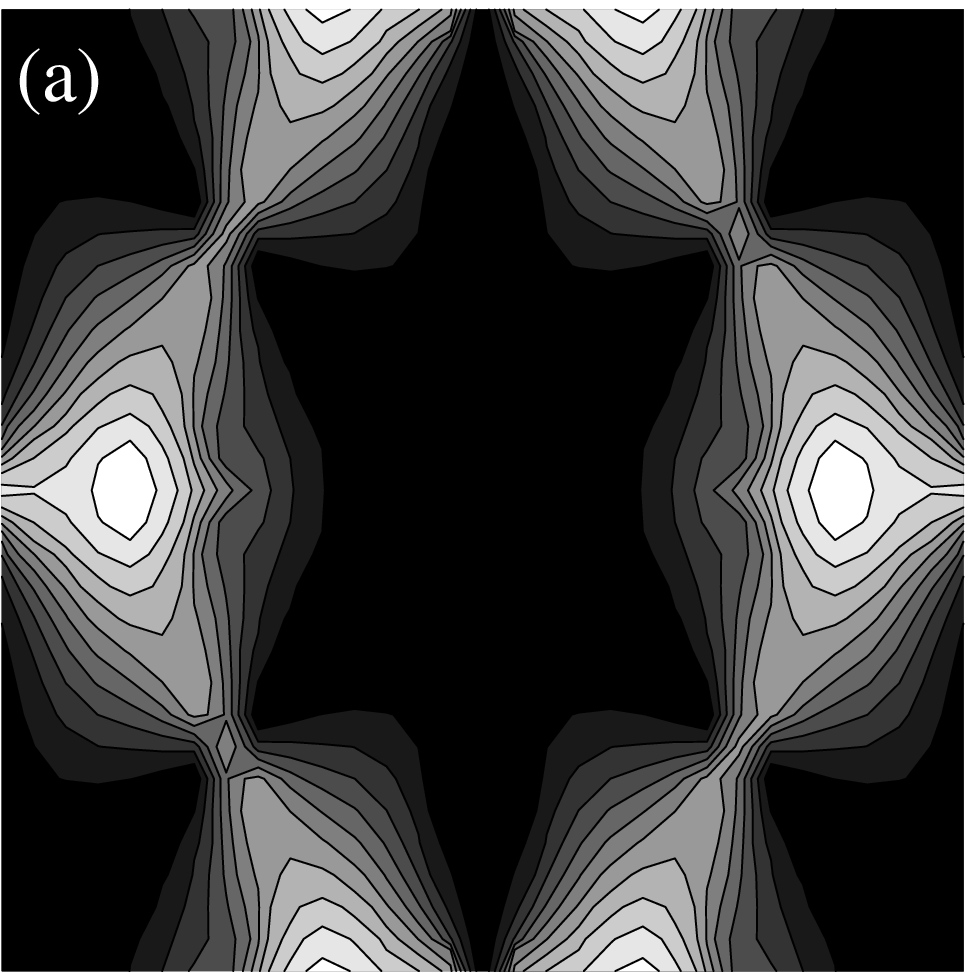}
\includegraphics[width=1.4in]{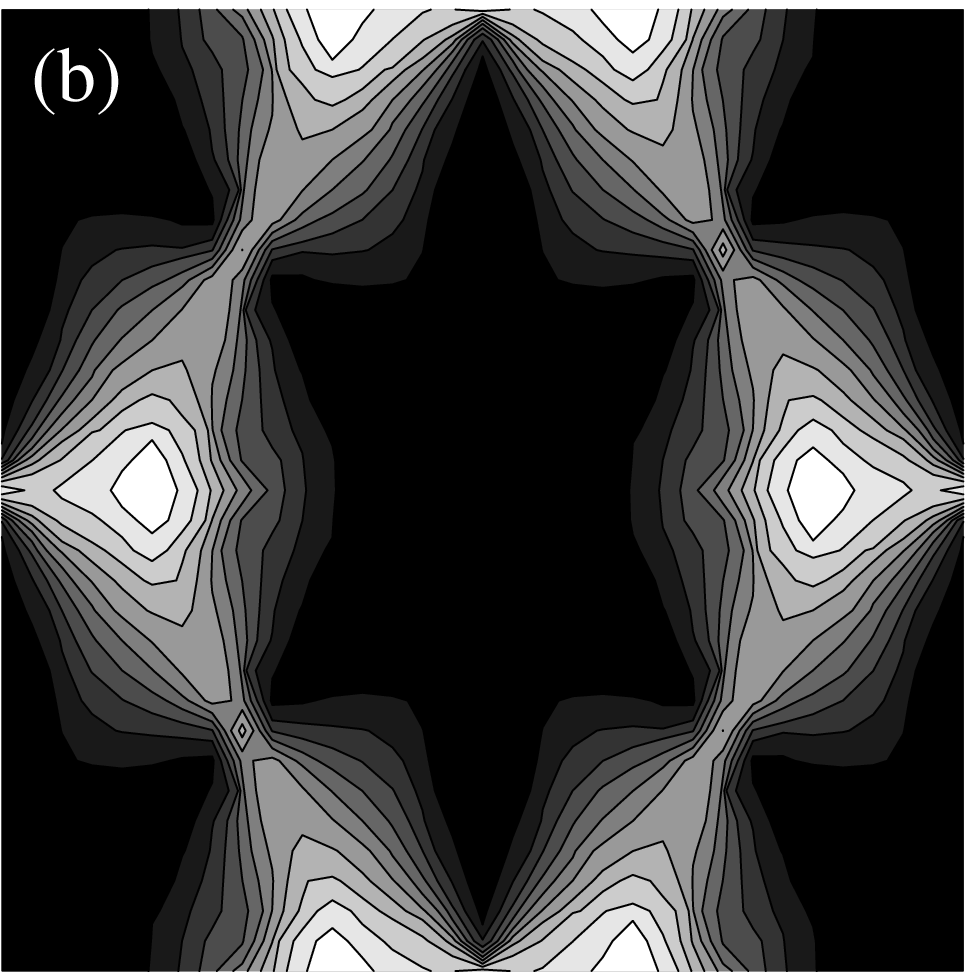}
\caption{ Contour plots of the structure factor in the
$[hhl]$ plane. (a) large-$N$ theory at zero temperature. (b)
Monte Carlo simulations at $T/J=1/100$ and $L=8$. Axes range from $-4\pi$ to
$4\pi$ and both plots are at the same resolution.
}
\label{sfplots}
\end{figure}

{\it Large-N mean field theory}.---Following Refs.~\onlinecite{garcan,isakov04},
we rewrite the Hamiltonian as
$H = {T \over 2} \sum_{i,j} M_{ij} {\bf S}_i \cdot {\bf S}_j$,
where $M_{ij}$ is the interaction matrix that has the information
about the nearest-neighbor interaction. The corresponding partition function is 
given by $Z = \int {\cal D} \bm{\phi} \ {\cal D} \lambda \ e^{-S(\bm{\phi},\lambda)}$
with the action 
$
S(\bm{\phi},\lambda) = \sum_{i,j}  \left [ {1 \over 2}
M_{ij} \bm{\phi}_i \cdot  \bm{\phi}_j +  {\lambda_i \over 2} \delta_{ij}
(\bm{\phi}_i \cdot \bm{\phi}_i - N) \right ],
$
where $\bm{\phi}_i$ 
is an $N$-component real 
vector field and $\lambda_i$ the Lagrange multiplier for the
constraint $\bm{\phi}_i \cdot \bm{\phi}_i = N$.

Now we take the $N \rightarrow \infty$ limit and set
a uniform $\lambda_i = \lambda_0$. 
The locations $i = (l,\mu)$ of spins can be labeled by 
those of the cubic unit cell $l = 1,..., n_c$ and the lattice sites 
$\mu=1,...,12$ within the unit cell ($n_c$ is the total number of the unit cells
in the lattice).
The Fourier transform with respect to the positions of the unit cells leads to
$S = \sum_{\bf q} \sum_{\mu,\nu} {1 \over 2}
A^{\mu \nu}_{\bf q} \bm{\phi}_{{\bf q},\mu} \cdot \bm{\phi}_{{\bf q},\nu}$
with $A^{\mu \nu}_{\bf q} = M^{\mu \nu}_{\bf q} + \delta_{\mu \nu} \lambda_0$.
Here $\lambda_0$ and  the eigenvalues, $m_{{\bf q},\rho}$, 
of the $12 \times 12$ 
interaction matrix $M^{\mu \nu}_{\bf q}$ are
determined by the saddle point equation,
$12 n_c = \sum_{\bf q} \sum_{\rho=1}^{12} 
{1 \over \lambda_0 + m_{{\bf q},\rho}}$.

It is found that the lowest eigenvalue is four-fold degenerate and 
independent of the wavevector.  The next lowest eigenvalue has a dispersion
and becomes the same as the lowest eigenvalue only at $q=0$. 
These features are very similar to those in the kagome and pyrochlore
lattices. These results imply that the spin structure of this system is
indeed highly frustrated and that magnetic order is suppressed.
The static spin-spin correlation function can be computed via
\cite{garcan,isakov04}
$\langle {\bf S}_{{\bf q},\mu} \cdot {\bf S}_{-{\bf q},\nu} \rangle
= \sum_{\rho = 1}^{12} {U_{{\bf q},\mu \rho} U_{-{\bf q}, \nu \rho} \over
\lambda_0 + m_{{\bf q},\rho}}$,
where $U_{{\bf q},\mu \rho}$ is a unitary transformation that diagonalizes
the interaction matrix, $M^{\mu \nu}_{\bf q}$. 
At zero temperature, the four degenerate eigenvalues dominate the
behavior of the spin-spin correlation function.
The resulting zero temperature structure factor,
$S({\bf q}) = \sum_{\mu \nu} \langle {\bf S}_{{\bf q},\mu} 
\cdot {\bf S}_{-{\bf q},\nu} \rangle$, in the [$hhl$] plane of the 
reciprocal space is shown in Fig.~\ref{sfplots}. The presence of high
intensity along bow-tie structures is apparent and qualitatively similar to that found
on the pyrochlore lattice. As discussed below, the structure factor in
the large-$N$ limit is very similar to that found by Monte Carlo
for the Heisenberg model ($N=3$) above the nematic ordering transition
temperature, but quite different from the Ising ($N=1$) case. 

\begin{figure}[t]
\includegraphics[width=1.3in]{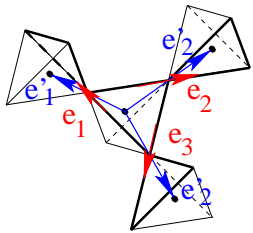}
\caption{ (color online). The dipolar (${\bf e}_{\kappa}$) 
and dual lattice (${\bf e'}_{\kappa}$) vectors.
}
\label{evecs}
\end{figure}

{\it Dipolar spin correlations}.---We found that the real space spin-spin
correlation function at long distances
is well described by the following dipolar form.
\begin{equation}
\langle S^{\alpha}_i S^{\beta}_j \rangle \propto  
\delta_{\alpha \beta} \left[
{3 ({\bf e}_i \cdot {\bf r}_{ij})({\bf e}_j \cdot {\bf r}_{ij}) \over |{\bf r}_{ij}|^5}
- {{\bf e}_i \cdot {\bf e}_j \over |{\bf r}_{ij}|^3} \right],
\label{dipcorr}
\end{equation}
where ${\bf r}_{ij}$ is a vector connecting sites $i$ and $j$,
and $\alpha,\beta=x,y,z$.
The ``dipolar vectors'' ${\bf e}_i$ are shown in Fig.~\ref{evecs}. 

In analogy to the pyrochlore \cite{isakov04}, 
we may understand the spin 
correlations in this system by mapping to a pure Maxwellian 
action with a ``Gauss law'' constraint. We first consider a dual lattice
of the hyperkagome lattice; the sites on the hyperkagome lattice
should be placed on the bonds of the dual lattice. This dual lattice
can be obtained by connecting centers of the tetrahedra of the
underlying pyrochlore lattice, but only along the directions
that would pass through the sites of the hyperkagome lattice.
Then there exists a unique bond $\kappa$ (on the dual
lattice) for a given site $i$. Let us define ${\bf e'}_{\kappa}$
as the unit vector along the bond $\kappa$ (see Fig.~\ref{evecs}).
We now define $N$ number of 
``magnetic'' vector fields ${\bf b}^{\alpha}_{\kappa}$ along the bond $\kappa$
via ${\bf b}^{\alpha}_{\kappa} = S^{\alpha}_{\kappa}  {\bf e}_{\kappa}$, 
where $S^{\alpha}_{\kappa}$
and ${\bf e}_{\kappa}$ represent the spin and unit vector 
defined earlier on the hyperkagome lattice.
Notice that the direction of these ``magnetic'' vector fields is not
along ${\bf e'}_{\kappa}$.
Nonetheless the ``magnetic'' vector fields 
satisfy ${\rm Div} \ {\bf b}^{\alpha} = \sum_{\kappa} 
{\bf e'}_{\kappa} \cdot {\bf b}^{\alpha}_{\kappa} =
\sum_{\kappa} {\bf e'}_{\kappa}
\cdot {\bf e}_{\kappa} S^{\alpha}_{\kappa} = 0$ on the dual lattice.
This is a direct consequence of the constraint
$\sum_{i \epsilon \Delta} S^{\alpha}_i = 0$ and 
${\bf e'}_{\kappa} \cdot {\bf e}_{\kappa} = \sqrt{2/3}$ for all $\kappa$.

We now define coarse grained
``magnetic'' vector fields, $B^{\alpha}$, averaged over clusters of spins \cite{isakov04,dipolar}.
There exist many ``flippable'' spin configurations where 
local rearrangement of the spins in a cluster can be made without 
violating the constraints. The coarse-grained field over such ``flippable'' spin 
configurations will average out to a small value. Then the large entropic
weight is related to the small values of $B^{\alpha}$. This feature
can be represented by an entropic weight of the form 
$\exp(-{K \over 2} \int d{\bf r} {\bf B}^2)$ \cite{isakov04,dipolar}.
This Maxwellian form of the ``action''
and the ``Gauss law'' constraint will lead to the dipolar form 
of $\langle B^{\alpha}_i ({\bf r}) B^{\beta}_j (0) \rangle \propto \delta_{\alpha \beta} (3 x_i x_j - r^2 \delta_{ij})/r^5$, 
and hence the spin-spin correlation function in Eq.~\ref{dipcorr}.
The discovery of such spin correlations supports the 
entropic argument {\it a posteriori}. 

\begin{figure}[t]
\includegraphics[width=2.8in]{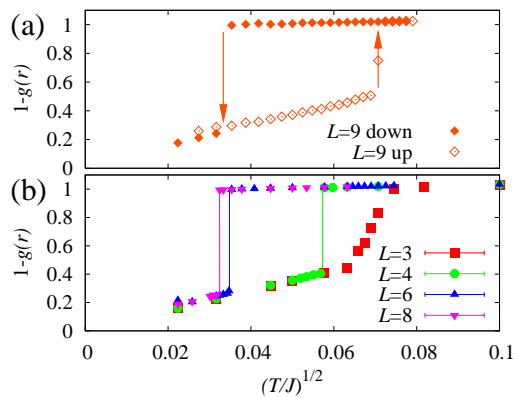}
\caption{ (color online). The nematic correlation function for 
the next-nearest neighbor triangles. (a) Hysteresis is observed upon lowering (down) or raising (up) the temperature; (b) finite size effects scale to a finite first order transition temperature (lines guide the eye).
}
\label{coplanar}
\end{figure}

\begin{figure}[t]
\includegraphics[width=2.8in]{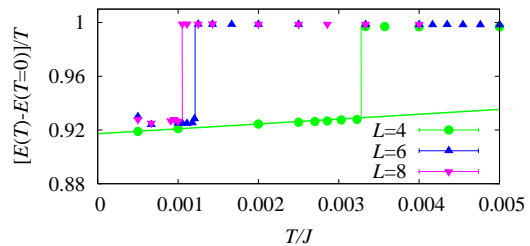}
\caption{ (color online).
The energy $E(T)-E(T=0)$ per spin divided by temperature. In the
$T\rightarrow0$ limit, this quantity approaches $11/12$ and equals the
specific heat.
}
\label{specific}
\end{figure}

{\it Monte Carlo simulations}.---Classical Monte Carlo simulations for
the Heisenberg ($N=3$) and Ising ($N=1$) models 
are performed on $L \times L \times L$ clusters of unit cells.
We mostly discuss the results of the Heisenberg model here
and mention those of the Ising model only as necessary.
In the first place, a first-order transition occurs in the Heisenberg model.
This can be most clearly seen in the nematic correlation function
defined as \cite{chalker92}
\begin{equation}
g({\bf r}_a-{\bf r}_b)=\frac{3}{2} \langle(
  {\bf n}_a \cdot {\bf n}_b)^2\rangle -\frac{1}{2},
\end{equation}
and
$
{\bf n}_a=\frac{2}{3\sqrt{3}}({\bf S}_1\times{\bf S}_2
  +{\bf S}_2\times{\bf S}_3+{\bf S}_3\times{\bf S}_1),
$
where ${\bf S}_1$, ${\bf S}_2$, and ${\bf S}_3$ are three spins on 
the triangle $a$. $g({\bf r})=1$ in an ideal coplanar state and 
$g({\bf r})=0$ in a non-coplanar state. The nematic correlation function for
the next-nearest neighbors is shown in Fig.~\ref{coplanar} and it clearly
shows a first order transition from a low temperature nematic ordered
state to a disordered state.  Hysteresis associated to this transition occurs 
in the temperature window $(1-5)\times 10^{-3}J$;
coplanar configurations are chosen below this window via
``order by disorder''. 
Similar behavior is seen for the nearest-neighbor and 
all higher-neighbor correlations.
The energy and specific heat data are also consistent
with the first order transition to nematic order (see Fig.~\ref{specific}).
Notice that the Monte Carlo data for the three largest 
system sizes $L=6,8$ and $9$ are almost identical. 
The zero temperature specific heat approaches 11/12 per spin 
which is consistent with the expectation that the low temperature
phase is dominated by coplanar spin structures. Analysis about
coplanar states tells us that there are 4 quartic and 20 quadratic 
modes per unit cell. Since each quartic (quadratic) mode 
contributes 1/4(1/2) to the specific heat \cite{chalker92}, 
the total specific heat becomes
11/12 per spin. Interestingly the same zero temperature specific heat was 
obtained in the Heisenberg model on the kagome lattice \cite{chalker92}.
The crucial difference between two cases, however, is that the
nematic order on the hyperkagome lattice is long-ranged at finite 
temperatures while it becomes long-ranged only in the $T \rightarrow 0$ limit 
on the kagome lattice \cite{chalker92}.

On the other hand, no magnetic order is seen prior to nematic order
as we do not find any elastic peaks in
the spin structure factor. We cannot, however, reliably comment whether
there is a magnetic ordering or not at still lower temperatures.
 
The spin correlations in the Heisenberg model in the disordered phase
are very similar to those in the large-N mean field theory 
(see Fig.~\ref{sfplots}) and can be fitted to the 
dipolar form in Eq.~\ref{dipcorr}.
On the other hand, the results on the Ising model are markedly different;
the spin-spin correlation function decays exponentially.
We expect this is due to the different form
of the local constraint in the Ising model. 


\begin{figure}[t]
\includegraphics[width=2.8in]{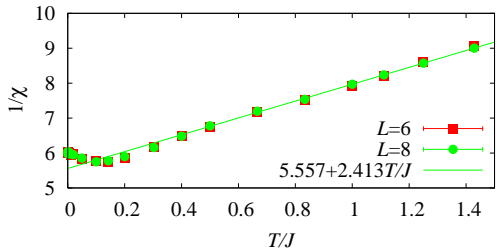}
\caption{ (color online). The inverse susceptibility as a function of
temperature, fit to a Curie-Weiss form for $T/J=0.9-2.0$.}
\label{sus}
\end{figure}

{\it Magnetization and susceptibility in an external field}.---The 
magnetization as a function of an external magnetic field, $h$, is
computed by the Monte Carlo simulations on the Heisenberg model. 
At finite temperatures, a weak plateau develops at $h/J = 2$, which
leads to a singular structure in the susceptibility (not shown).
This can be explained by the occurrence of a collinear 
order (up-up-down spin structure) by disorder at $h/J = 2$
in analogy to the kagome lattice case \cite{zhit}.

{\it Implications for experiments}.---A Curie-Weiss fit to the Monte 
Carlo susceptibility data for $T \gtrsim J$, leads to
$\theta_{CW} = - 2.303(5)J$ (see Fig.~\ref{sus}). Comparing this with
the experimental value $\theta_{CW} = -650K$ \cite{takagi},
one obtains $J \approx 280 K$.
This suggests that the nematic 
transition may occur around  $0.3-1.5 K$ if our results
are taken seriously, and below this temperature coplanar spin configurations 
would be preferred. 
Even though our classical computations may not be directly
applicable at such low temperatures,
we suspect that coplanar spin configurations
may still dominate at low temperatures even in the quantum regime
\cite{comment}.

Our results also suggest that the spin correlations at
$T > J/100 \sim$ 2-3$K$ may be dominated by the physics of the classical 
spin liquid with dipolar spin correlations; this will be checked by
neutron scattering experiments. Notice that there is no sign of 
magnetic ordering down to 2-3$K$ in the experiment \cite{takagi}; 
this may also be consistent with our Monte Carlo results that
show no evidence of magnetic ordering above the nematic transition.
It remains to be seen whether the system develops
 magnetic order at very low temperatures by ``order by disorder''
due to classical/quantum fluctuations or prefers a magnetically-disordered 
quantum spin liquid. Finally, in the current work, we have not considered 
the orbital degree of freedom
which may play an important role in the real material. Studies of
quantum spin models and the role of orbital degrees of freedom,
therefore, are important subjects of future study.
 
{\it Acknowledgments}: We are grateful to Hide Takagi for showing us
his unpublished experimental data. We thank A. Paramekanti,
M. Lawler, and especially A. Vishwanath for helpful discussions. 
This work was supported by the NSERC, CRC, CIAR
(JMH, SVI, HYK, YBK), the Sloan Fellowship (HYK),
KRF-2005-070-C00044 and the Miller Institute of Basic 
Research in Science at UC Berkeley via the Visiting Miller 
Professorship (YBK).


\begin{thebibliography}{99}
\bibitem{comment}
At this time of writing, the experimental data below 2-3$K$ is not available.
\end{thebibliography}
\end{document}